\documentclass[]{aa}
\usepackage{natbib}
\usepackage{graphicx}
\usepackage{wasysym}
\usepackage{txfonts}
\usepackage{color}
\def\rmv{\mbox{\rm v}}
\def\rmdv{\mbox{\rm dv}}
\def\fH2{\mbox{f$_\HH$}}

\def\EBV{\mbox{E$_{\rm B-V}$}}

\def\nH2{\mbox{${\rm n}_\HH}$}
\def\Rgal{\mbox{R$_{\rm gal}$}}
\def\pccc{~{\rm cm}^{-3}} 
 
\def\pcc{~{\rm cm}^{-2}}

\def\Tsub#1 {\mbox{${\rm T}_{\rm #1}$}}
\def\TK  {\Tsub K }

\def\Tsp {\Tsub sp }

 \def\arcmin{\mbox{$^{\prime}$}}

\def\degr{$^{\rm o}$}
\def\p{\mbox{$^+$}}

\def\cch{\mbox{C$_2$H}}

\def\h13cop{\mbox{{H$^{13}$CO\p}}}

\def\C3H{\mbox{C$_3$H}}
\def\c3h2{\mbox{C$_3$H$_2$}}
\def\cc3h2{\mbox{{\it c}-C$_3$H$_2$}}
 
 \def\R0{R$_0$}
\def\G0{\mbox{G$_0$}}

\def\ddeg{{}^\circ\kern-.1em}

\def \kms{\mbox{km\,s$^{-1}$}}

\def\E#1 {$10^{#1}$}
\def\E#1 {E{#1}}
\def\P#1,{$\nH2\TK~=~#1\times~10^4\pccc$~K}
\def\ec#1,#2,#3,{#1\,(#2)\E{#3}}

\def\H3{\mbox{H$_3$}}

\def\RH2{\mbox{R$_{\rm G}$}}
\def\g13{\mbox{g$_{13}$}} 

\def\cc3h{\mbox{{\it c}-\C3H}}
\def\lc3h{\mbox{{\it l}-\C3H}}

\newcommand{\emm}[1]{\ensuremath{#1}}   
\newcommand{\emr}[1]{\emm{\mathrm{#1}}} 

\newcommand{\hcop}{\emr{HCO^+}} 
 
\newcommand{\HH}{\emr{H_2}}

\renewcommand{\coth}{\emr{^{13}CO}}

\newcommand{\Msun}{\emm{M_\odot}}

\title{ALMA observations of molecular absorption in four directions toward the Galactic bulge}

\author{ H. Liszt\inst{1} and M. Gerin\inst{2}}
\institute{
     National Radio Astronomy Observatory,
           520 Edgemont Road,
           Charlottesville, VA,
           USA 22903 
      \email{hliszt@nrao.edu}
\and
LERMA, Observatoire de Paris,  PSL Research University, CNRS,
Sorbonne Universit\'es, UPMC Univ. Paris 06, Ecole Normale
  Sup\'erieure, F-75005 Paris, France
\email{maryvonne.gerin@ens.fr}
}
\begin{document}
\date{received \today}%
\offprints{H. S. Liszt}%
\mail{hliszt@nrao.edu}%
%
\abstract
{Alma Cycle 3 observations serendipitously showed strong absorption from
diffuse molecular gas in the Galactic bulge at 
-200 \kms\ $< {\rm v} < -140$ \kms\ toward the compact extragalactic
continuum source J1744-3116 at\ (l,b)= -2.13\degr,-1.00\degr. }
{We aimed to test whether  molecular gas in the bulge could also be detected toward the three
other, sufficiently strong  mm-wave continuum sources seen toward the bulge
at $|b| < 3$\degr.}
{We took absorption profiles of \hcop (1-0), HCN(1-0), \cch (1-0), CS(2-1) 
and H$^{13}$CO\p (1-0) in 
 ALMA Cycle 4 toward J1713-3418, J1717-3341, J1733-3722 and J1744-3116.}
 {Strong molecular absorption from disk gas at $|\rmv| \la 30$ \kms\ was detected 
  in all directions, and absorption from the 3 kpc arm was newly detected toward J1717
  and J1744.  However, only the sightline toward J1744 is dominated by molecular 
 gas overall and no other sightlines showed molecular absorption from gas deep inside 
 the bulge.  No molecular absorption was detected toward  J1717 where H I emission from 
 the bulge was previously known.  As observed in \hcop, HCN, \cch\ and CS, the bulge gas toward 
 J1744 at $v < -135$ \kms\ has chemistry and kinematics like that seen near the Sun and
  in the Milky Way disk generally.  We measured isotopologic
  ratios N(\hcop)/N(H$^{13}$CO\p) $> 51~(3\sigma)$ for the bulge gas toward J1744
  and $58\pm9$ and $64\pm4$ for the disk gas toward J1717 and J1744, 
 respectively, all well above the value of 20-25 typical of the central molecular zone.}
 {The kinematics and chemistry of the bulge gas observed toward J1744 more
 nearly resemble that of gas in the Milky Way disk than  in the central 
  molecular zone.}

\keywords{ interstellar medium -- abundances; B1741-312 }

\authorrunning{Liszt,Gerin} \titlerunning{More on diffuse molecular gas in the Galactic bulge}

\maketitle{}

%

\section{Introduction}

In a recent paper \citep{GerLis17} we discussed the serendipitous discovery of 
strong molecular absorption at -210 \kms\ $\le$ v $\le$ -135 \kms\ 
toward the compact extragalactic continuum source J1744-3116 (aka B1741-312) 
at $l,b$ = -2.13\degr,-1.00\degr,  along a line of 
sight passing through the inner portions of the Galactic bulge outside the 
central molecular zone (CMZ; \cite{MorSer96}).  We extracted the run of 
IR K-band extinction with distance modulus from the 3-D extinction maps 
of \cite{SchChe+14} and showed that the K-band extinction in the bulge 
implied the existence of unseen gas, given the weakness of
H I emission at high negative velocities characteristic of the bulge.
The \HH\ column density inferred from \hcop\ absorption matched the
amount of gas lacking in H I.

As we discussed, the quantity of molecular gas in the bulge outside the CMZ, 
at 1.2\degr\ $\la |l| \la 10$\degr\ or galactocentric radius 
180 pc $\la$ R $\la$ 1500 kpc at the IAU standard distance R$_0$ = 8.5 kpc, 
is rather uncertain.  The total mass of H I in the bulge is comparable to
the mass of \HH\ in the CMZ, M(H I) $\approx 3\times 10^7$ \Msun \citep{BurLis78},
raising the possibility that the mass of molecular gas in the bulge is yet
larger.

Mm-wave absorption toward J1744 from a wider range of species was subsequently 
discussed 
by \cite{RiqBro+17} who pointed out some of the same similarities between the 
chemical composition of the bulge and disk gases that are noted here.  
In the present work we re-observed molecular absorption toward J1744 with
more complete spectral coverage than in our earlier discussion, at much 
higher spectral resolution than in our earlier work or in \cite{RiqBro+17}.
We also searched for similar absorption toward the few (three) known, 
strong mm-wave continuum sources lying toward the bulge within 3\degr\ of the 
Galactic plane.  All of the newly-observed sightlines lie much further from 
the Galactic center than J1744, and none of them show molecular absorption 
at velocities characteristic of gas deep inside the bulge: we did find 
two weak features arising in the 3 kpc arm \citep{Ban77,Ban80,DamTha08}.

The plan of this work is as follows: in Sect. 2 we describe the new and
existing observational material that is discussed.  In Sect. 3 we briefly
summarize some properties of the observed sightlines and the disk gas 
that is present along them.  In Sect. 4 we present new observational 
results leading to a better characterization of the bulge gas toward 
J1744: it has a chemistry and cloud structure comparable to that seen 
in \hcop, \cch, HCN and CS in diffuse molecular gas near the Sun and, 
unlike material in the CMZ, is not enhanced in $^{13}$C.  Section 5 
is a summary and discussion.

\section{Observations and data reduction}

\begin{figure*}
\includegraphics[height=14cm]{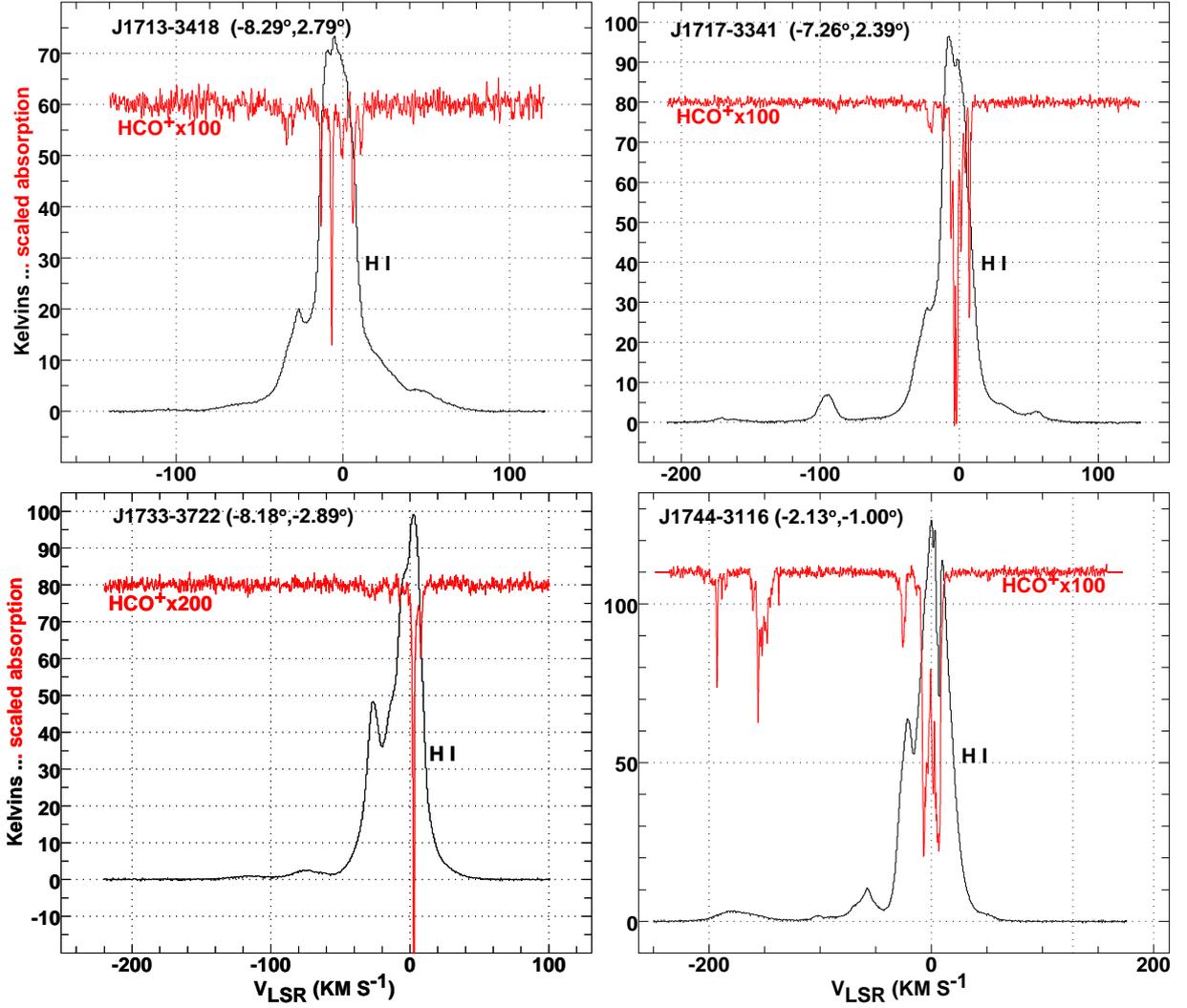}
  \caption[]{GASS H I and ALMA J=1-0 \hcop\ profiles for the four sources 
observed here with ALMA.}
\end{figure*}

\begin{figure*}
\includegraphics[height=15cm]{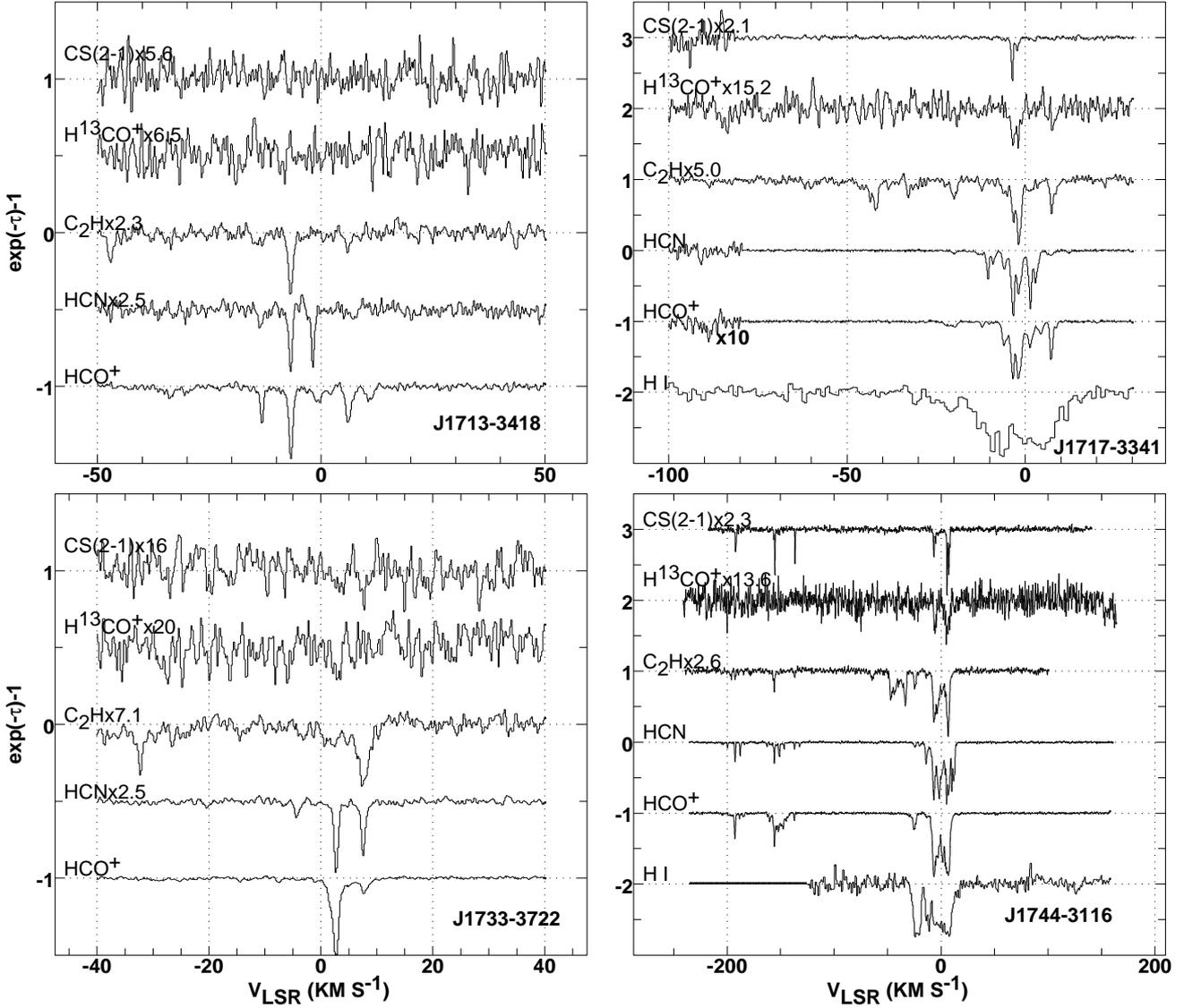}
  \caption[]{ALMA absorption profiles for the four sources observed here.}
\end{figure*}

\subsection{New ALMA absorption measurements}

We observed the J=1-0 lines of \hcop, H$^{13}$CO\p, HCN and \cch, and the J=2-1 line
of CS J=2-1 in absorption toward the four continuum sources listed in Table 1. The
work was conducted  under ALMA Cycle 4 project 2016.1.00132.S whose pipeline data 
products were delivered in 2017 February. The spectra discussed here were 
extracted from the continuum-subtracted pipeline-processed data cubes at the 
pixel of peak continuum flux in the continuum map
made from each spectral window, and divided by the continuum flux in the
continuum map at that pixel. Fluxes at 89.2 GHz ranged from 0.33 Jy for J1713 to 
0.93 Jy for J1733.  Each spectrum consisted of 1919 semi-independent channels
spaced 61.13 kHz corresponding to 0.205 \kms\ at the 89.189 GHz rest frequency of \hcop\
or 0.187 \kms\ for CS(2-1) at 97.981 GHz: the channel spacing is half the
spectral resolution.  Velocities presented with the spectra are
taken with respect to the kinematic definition of the Local Standard of Rest.

\subsection{Other data}

In Figure 1 we compare our \hcop\ spectra with Galactic All-Sky Survey (GASS) III 
$\lambda$21 cm H I emission spectra \citep{KalHau15}. For the sources J1717 and 
J1744 we use the H I absorption profiles of \cite{DicKul+83} to derive an 
optical depth-corrected column density of atomic hydrogen as

$$N(H~I) = 1.823\times 10^{18}\pcc \int \frac{\tau(\rmv)T_B(\rmv)}{1-\exp{(-\tau(\rmv))}} \rmdv,
\eqno(1) $$

where $\tau$ and T$_B$ are the optical depth and brightness temperature
and the units of velocity are \kms.   In this work, N(H) is the column 
density of H-nuclei in neutral atomic and molecular form,N(H) = N(H I)+2N(\HH).

We also use the CS and HCN absorption profiles for local clouds from
\cite{LucLis02} and \cite{LisLuc01}, respectively, and the \hcop\ and
\cch\ measurements of \cite{LucLis00C2H}.

\subsection{Extinction}

We cite the 6\arcmin\ resolution dust-emission maps scaled to optical
reddening \EBV\ by \cite{SchFin11} and the 3-D IR extinction map
 of the Galactic bulge of \cite{SchChe+14} at the same resolution, 
from which we cite values of the K-band extinction A$_{\rm K}$.

\section{Disk gas toward the 4 targets}

Profiles toward all sources are shown in Figs. 1-3 and some general properties of 
the sightlines are given in Table 1.  All directions show strong molecular
absorption from gas in the Galactic disk around zero velocity but only the sightline 
to J1744 passes inside galactocentric radius R = 1 kpc and only that sightline 
shows strong molecular absorption from gas deep inside the Galactic bulge.

The sightline to J1744 is also the only one for which the gas column density is 
dominated by molecular gas, mostly from the disk  contribution around
0-velocity.  The species-by-species chemistry of this gas is not a major
focus of this work, but is shown in Sect. 4 along with a variety of
other data that are used as a basis of comparison with the bulge gas. 
Chemistry of the disk gas toward J1744 was discussed for a broader range 
of species observed in a low spectral resolution spectral scan by 
\cite{RiqBro+17}, whose conclusions agree with ours regarding the 
similarity of the disk and bulge diffuse molecular gas components.

Although some of the molecular gas  must be quite close to the Sun, for
instance material in the Heeschen Cloud \citep{Hee55} that is responsible 
for the H I self-absorption toward J1744, 
it is hard to place the disk gas in Galactic perspective .
The line of sight velocity gradients due to Galactic rotation are quite
shallow at the small longitudes of our sources: for a flat rotation curve
with $\Theta(R) = 220$ \kms, a location 5 kpc from the Sun has v $\approx -10$ 
\kms\ toward J1744 or v $\approx  -40$ \kms\ toward the other sources.  
Moreover, much of the low-velocity  absorption occurs at positive velocities 
that are inconsistent with rotation. Positive velocities could arise from a 
combination of random motions and streaming associated with spiral arms, 
although gas at v $\ga$ 30 \kms\ most likely arises inside the Galactic disk.

The \HH\ fraction falls monotonically with separation from the Galactic equator:
only the line of sight toward J1744 at b=-1\degr\ has a molecular fraction exceeding 50\%.
The optical depth corrections to N(H I) in the two directions with measured
H I absorption profiles are 35\%-45\% of the optically thin value, but these 
sightlines have relatively
high \HH\ fractions and the correction to N(H I) is smaller than 2N(\HH).

\subsection{$^{12}$C/$^{13}$C in the disk gas}

Profiles of the carbon isotopologues of \hcop\ are shown in Fig. 3 and results 
for their profile integrals and ratios of column densities are given in Table 2.
Column densities are derived from integrated optical depths in the limit of 
no rotational excitation above the cosmic microwave background, for
a molecular dipole moment of 3.92 Debye (eg \cite{AndKoh+16}).
From this it follows that N(\hcop) = $1.10\times 10^{12} \pcc \int \tau~\rmdv$ 
and N(H$^{13}$CO\p) $= 1.135\times 10^{12} \pcc \int \tau~\rmdv$ in units 
of \kms,  accounting for the lower line frequency of the heavier isotopologue.  
Error estimates for the integrated 
optical depths in Table 2 include effects associated with the appreciable optical 
depths in \hcop, which were estimated using the measured rms line/continuum ratios 
in absorption-free regions in channel-by-channel Monte Carlo simulations of the 
observed profile taken as the expectation value.

We have precise measurements of the carbon \hcop\ isotopologues toward two 
sources: N(\hcop)/N(H$^{13}$CO\p) $= 58 \pm 9$ toward J1717 and
N(\hcop)/N(H$^{13}$CO\p) $= 64 \pm 4$ toward J1744.  This is well below the 
Solar isotopic abundance ratio 89, but typical for gas near the Sun 
observed in mm-wave absorption by \cite{LucLis98}. However, some of the 
disk gas observed here could be situated well inside the Solar Circle 
where the $^{12}$C/$^{13}$C ratio is known to be smaller \citep{MilSav+05},
although still well above the very small values 20 - 25 characteristic 
of the CMZ \citep{Wil99,RiqAmo+10}.

\subsection{H I spin temperature in the self-absorbed gas toward J1744}

The line of sight to J1744 is the exceedingly  rare case where H I 
absorption \citep{DicKul+83} has been measured in a direction where 
the H I emission is self-absorbed. Earlier \citep{GerLis17} we showed 
that the H I self-absorption is coincident with CO J=1-0 emission.
Here we show how the combination of H I absorption and self-absorption 
can be used to derive an upper limit on the spin temperature of the 
H I self-absorbing but largely molecular gas.

We define T$_{\ rm B}$ as the observed brightness temperature of the 
H I emission and T$_{{\rm B},{bg}}$ as the hypothetical brightness 
temperature in the absence of self-absorption: extrapolating across the 
self-absorption feature by Gaussian fitting of the H I emission gives 
T$_{{\rm B},{bg}}$ =  125.2 K in the center of the H I 
self-absorption trough where T$_{\rm B}$ = 71.1 K.  We also define the optical 
depth derived from the absorption profile of \cite{DicKul+83} as the 
sum of  foreground and background contributions $\tau = \tau_{fg}+\tau_{bg}$ 
and define T$_{{\rm sp},{fg}}$ and T$_{{\rm sp},{bg}}$ as the spin temperatures of
the foreground and background emitting gases, respectively.  Then
the two equations

$$ T_{{\rm B},{bg}}~\exp(-\tau_{fg}) + T_{{\rm sp},{fg}} (1-\exp(-\tau_{fg})) = T_B
\eqno(2) $$

and

$$ T_{{\rm B},{bg}} = T_{{\rm sp},{bg}} (1-\exp(-\tau_{bg})) \eqno(3) $$

can be solved for the spin temperatures of the foreground and background 
gases as functions of the value of $\tau_{fg}$. The minimum optical 
depth of the foreground gas is $\tau_{fg} = 0.58$ when T$_{sp,fg} = 2.73$ K.
This can be compared with the total optical depth in the middle of the
H I self-absorption trough, $\tau = 1.22$, and it implies a fairly high 
minimum spin temperature for the background gas seen in emission: 
$\tau_{bg} \le 1.22-0.58 = 0.64$ and T$_{{\rm sp},{bg}} \ge 265$ K.

The implied spin temperature of the foreground self-absorbing gas 
increases as $\tau_{fg}$ increases above 0.58 and the maximum 
optical depth, $\tau_{fg} = \tau = 1.22$, 
corresponds to \Tsp\ = 48K that is typical of gas observed in H I absorption 
generally \citep{HeiTro03}. Of course it is 
not realistic to assign all of the optical depth to any one component of
the H I and an improved upper bound on the spin temperature of the 
self-absorbing gas follows from showing that the optical depth
of the self-absorbing gas is below 1.22.  To this
end we decomposed the H I absorption profile of \cite{DicKul+83}
as summarized in Table 3 and illustrated in Fig. 4. A kinematic
component is found coincident with the self-absorption, having
a center optical depth $\tau_{fg} = 0.686\pm0.081$ corresponding to
${\rm T}_{sp,fg} = 16.2(+8.0,-10.2)$ K.  This result unfortunately 
spans the range of kinetic temperatures from dark (10 K) to diffuse (25 K) 
molecular gas and does not allow us to decide in which regime 
the foreground gas exists.

The precision of the present result is limited by the noise level in the 
rather old absorption profile of \cite{DicKul+83}. This could be improved 
by taking a more sensitive and contemporary H I absorption profile toward 
J1744, and extended by taking H I absorption profiles toward other positions 
with apparent H I self-absorption, which is relatively common in sightlines 
through the Great Rift in the inner Galaxy.  Further insight into the 
foreground gas toward J1744 could also be obtained by observing \coth\ in 
emission, and by observing CO or \coth\ in absorption.


\section{Gas inside the Galactic disk}

This work shows more clearly the nuclear bulge gas at v $< -135$ \kms\
toward J1744 that was discovered in our earlier work \citep{GerLis17}
(see also \cite{RiqBro+17}) 
but otherwise we did not detect gas inside the Galactic disk 
except for weak features arising in the 3 kpc arm at -90 \kms\ toward 
J1717 (Fig. 1) and at v = $+52$ \kms\ toward J1744.

\subsection{Cloud structure and chemistry in the bulge gas toward J1744}

\subsubsection{Strucure and kinematics}

As discussed in \cite{GerLis17}, especially their Table 1 and Fig. 4, 
gas at v $\la -135$ toward J1744 
arises in the Galactic bulge at galactocentric radii 
320 pc $\le$ \Rgal\ $\la 1500$ pc where 320 pc is the distance of
closest approach to the center and the outer radius depends on the 
contribution of non-circular motion to the observed velocity.  A purely 
circular description of the motion would place all the gas at 
\Rgal\ $\la 450$ pc.

No matter what velocity field is adopted, the line of sight velocity 
gradient is very steep  and the $\approx$ 20-25 \kms\ velocity intervals over which 
the molecular gas is seen, -205 $\la$ v $\la $-185 \kms\ and 
 -160 $\la$ v $\la $-135 \kms, correspond to only 100-150 pc along
the line of sight.  In each of these intervals, \hcop\ absorption
more nearly fills the velocity range (Fig. 5), analogous to what is seen in
gas near the Sun: local \hcop\ absorption is more ubiquitous than
absorption in HCN or CS and the column densities of HCN and CS increase
substantially for individual features having N(\hcop) $\ga 10^{12}\pcc$ or 
W$_{\hcop} \ga 1$ \kms\ \citep{LisLuc01,LucLis02}. 

A closer view of the absorption in the bulge gas is presented
in Fig. 5.  This figure  shows that absorption from CS and HCN is almost 
entirely concentrated in 
narrow-lined features having profile FWHM $\la 1$ \kms.  \hcop\ also 
shows these (and other) narrow features, although in \hcop\ they are superposed 
over a broad and presumably more nearly volume-filling component that 
also appears weakly in HCN at v $>$ -160 \kms.   

The appearance of so many narrow features, and so much kinematic 
substructure in the broader \hcop\ absorption is uncharacteristic of gas 
observed at small galactocentric radii. It likely arises because of the u
nusual circumstances 
under which the gas is being observed.  The extremely large line
of sight velocity gradient can kinematically separate gas parcels 
that are near each other in space and have little relative motion
with respect to each other.  In this way, spatial substructure is 
revealed that would otherwise remain hidden. If seen locally the 
same gas distribution would be narrower in velocity 
and the substructure seen toward J1744 would be blended together.

\subsubsection{Chemical comparisons in the bulge gas}

To illustrate the chemistry of the
bulge gas, comparing HCN and CS column densities, we decomposed the line
 profiles of HCN and CS into gaussian components as shown in Tables 4-5. 
To account for the HCN components that are not apparent in CS (the feature 
at -151.8 \kms\ and the broad line at -155.5 \kms) we subtracted the 
gaussian fits in the CS profile and integrated the remainder CS profile over the 
ranges of the HCN components.

The results are presented in terms of HCN and CS column densities in Fig.
6 at left where we also include results for the local gas seen earlier
at high Galactic latitude \citep{LisLuc01,LucLis02} and for the disk 
gas toward the four sightlines studied in this work.  In fact, the HCN and 
CS column densities and their ratios in the bulge gas are very similar to 
those seen in local gas, with the exception of a high CS/HCN ratio in the 
feature at -137 \kms.  Comparably high CS/HCN and CS/\hcop\ ratios are 
occasionally inferred from unpublished emission measurements we have made 
in the vicinity of the continuum background targets used by \cite{LisLuc01} 
and \cite{LucLis02} to study local gas in absorption.  CS absorption is 
comparatively weak in the Galactic disk gas toward J1733, and at
v $<$ 0 \kms\ toward J1744.  Similar conclusions regarding the similarity
of disk and bulge gas are evident in the figures of \cite{RiqBro+17}

The hydrocarbon chemistry is illustrated at right in Fig. 6 where we compare
\cch\ and \hcop\ in the new data with that shown earlier by 
\cite{LucLis00C2H}.  The bulge gas toward J1744 is at the lower edge of
the range of \cch/\hcop\ abundance ratios but still within a factor two
of the regression line through the older data. 

\subsection{$^{12}$C/$^{13}$C in the bulge gas seen toward J1744}

\hcop\ and H$^{13}$CO\p\ profiles are shown in more detail for three sources 
in Fig. 3. The profiles toward J1744 are shown in two parts corresponding to gas 
in the nuclear bulge at v $< -135$ \kms\ and Galactic disk material at 
v $>$ -40 \kms, separately scaled. Fig. 3 shows various velocity intervals 
over which the profiles were integrated to determine column densities, 
as given in Table 2 . 

For the gas in the nuclear bulge observed at v $< -135 $ \kms\ toward J1744 
(Table 2) we have only a weak limit $^{12}$C/$^{13}$C $>$ 17 for the sub-interval  
v $<$ -190 \kms, but $^{12}$C/$^{13}$C $>$ 51 overall and separately over the 
sub-interval $-165 \le \rmv \le -135$ \kms.  Apparently, this gas is not 
enhanced in $^{13}$C, which is very different from the values 
$^{12}$C/$^{13}$C $\approx$ 20-25 seen in the CMZ \citep{Wil99,RiqAmo+10}. This could be
disk gas that entered the bulge recently, or gas resident in the bulge that
has never been astrated beyond the values characteristic of the disk in the 
inner Galaxy.

\subsection{Inner-Galaxy gas along the line of sight to J1717}

As shown in Fig. 1, H I emission from the bulge is present toward
J1717 at -200 \kms\ $\la$ v $\la$  -140 \kms: this was modelled as part of 
the tilted H I inner-galaxy gas distribution by \cite{BurLis78}.  The H I 
column density N(H I) $= 4.6\times 10^{19}\pcc$ is comparable to that seen
toward J1744 at -165 \kms\ $\la$ v -135 $\la$ \kms, N(H I) 
$= 6.1\times 10^{19}\pcc$ \citep{GerLis17}, but the bulge gas toward J1717 
is not dominated by \HH: the $3\sigma$ limit on 2N(\HH) found by integrating 
the \hcop\ profile is 2N(\HH) $\le 5.0\times 10^{19}\pcc$.

As also shown in Fig. 1, a strong H I feature corresponding to the 3 kpc arm
\citep{Ban77,Ban80,DamTha08}
is present toward J1717 at v $\approx -90$ \kms.  \hcop\ absorption is also
present toward J1717 over the range -125 \kms\ $\la$ v -75 $\la$ \kms\  but
the gas is only weakly molecular:  we find N(H I) $= 2.0\times 10^{20}\pcc$ 
and 2N(\HH) $= 6.4 \pm 1.8\times 10^{19}\pcc$, or \fH2\ $\approx 0.24$.  

\subsection{The far-side 3 kpc arm toward J1744}

As seen in Fig. 1 there is a weak wing of H I emission at v $\ga 40$ \kms\
toward J1744, and weak \hcop\ absorption as well.  Integrating over the
velocity range 40-60 \kms\ we find N(H I) $= 8.0\times10^{19}\pcc$, 
2N(\HH) $= 7.3 \pm 1.2 \times 10^{19}\pcc$, ie, the gas is $\approx$ 48\%
molecular.  

\begin{figure*}
\includegraphics[height=15cm,angle=-90]{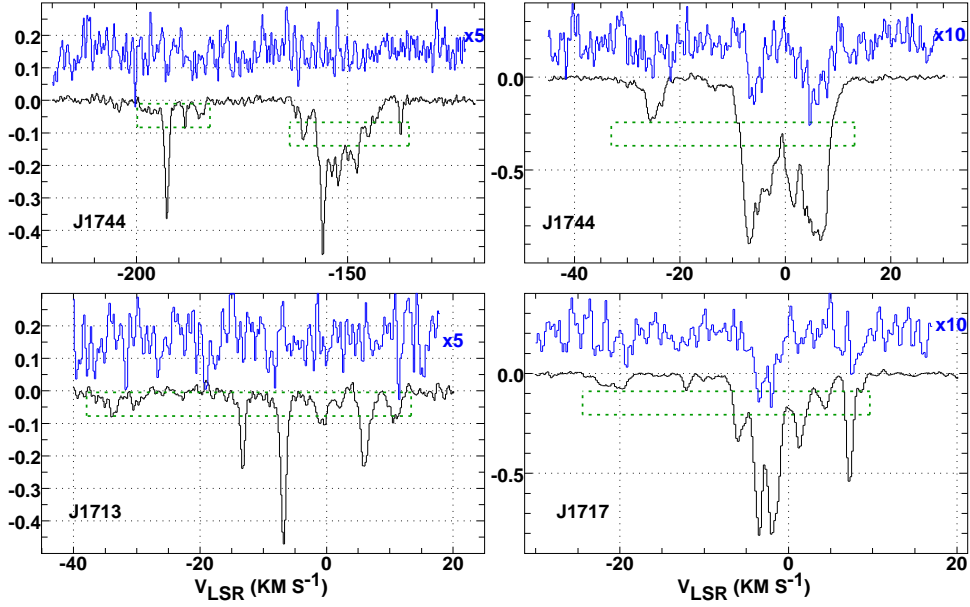}
\caption{\hcop\ and  H$^{13}$CO\p\ absorption profiles for three
of the sources observed here. The  H$^{13}$CO\p\ are scaled upward
as indicated and shown in blue over those of \hcop\. Absorption toward 
J1744 is shown in two parts in the upper panels, with different scaling.  
Regions over which the profiles were integrated are shown outlined, at 
the mean level within each region:  see Table 2 for results.}
\end{figure*}

\begin{figure}
\includegraphics[height=9.3cm]{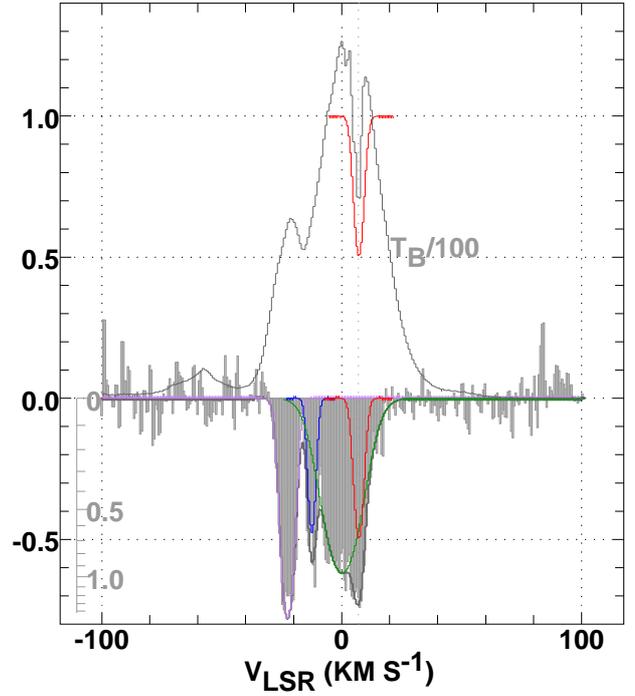}
\caption{$\lambda$21 cm H I emission, absorption and self-absorption 
toward J1744.  Bottom: Gaussian decomposition of the H I absorption profile 
of \cite{DicKul+83} (see Table 3).  An optical depth scale is shown inset.  
Top: GASS III H I emission scaled down by a factor 100, with one component 
of the Gaussian decomposition shown superposed to show coincidence between 
the H I absorption and self-absorption.}
\end{figure}

\begin{figure}
\includegraphics[height=8.8cm]{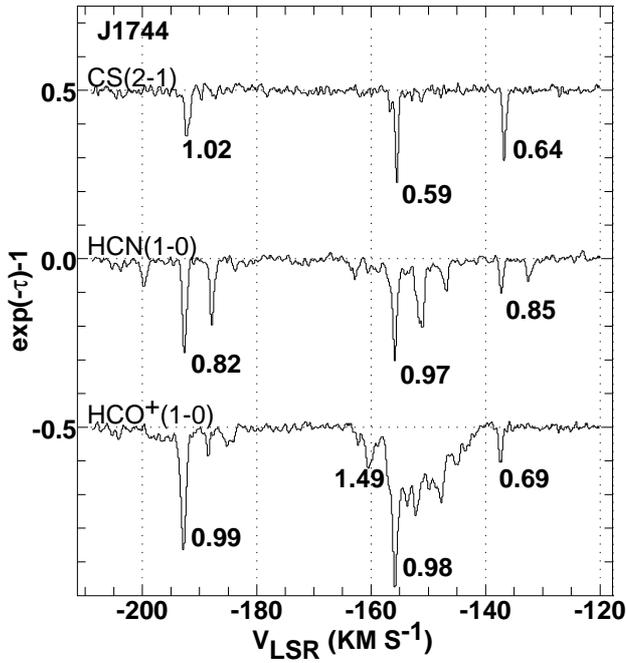}
\caption{\hcop, HCN and CS absorption profiles toward J1744 at velocities
arising inside the Galactic bulge.  The FWHM of gaussian fits to the narrow 
kinematic components are indicated in units of \kms: typical errors are 0.04 \kms.
See Tables 4-5.}
\end{figure}

\begin{figure*}
\includegraphics[height=7cm]{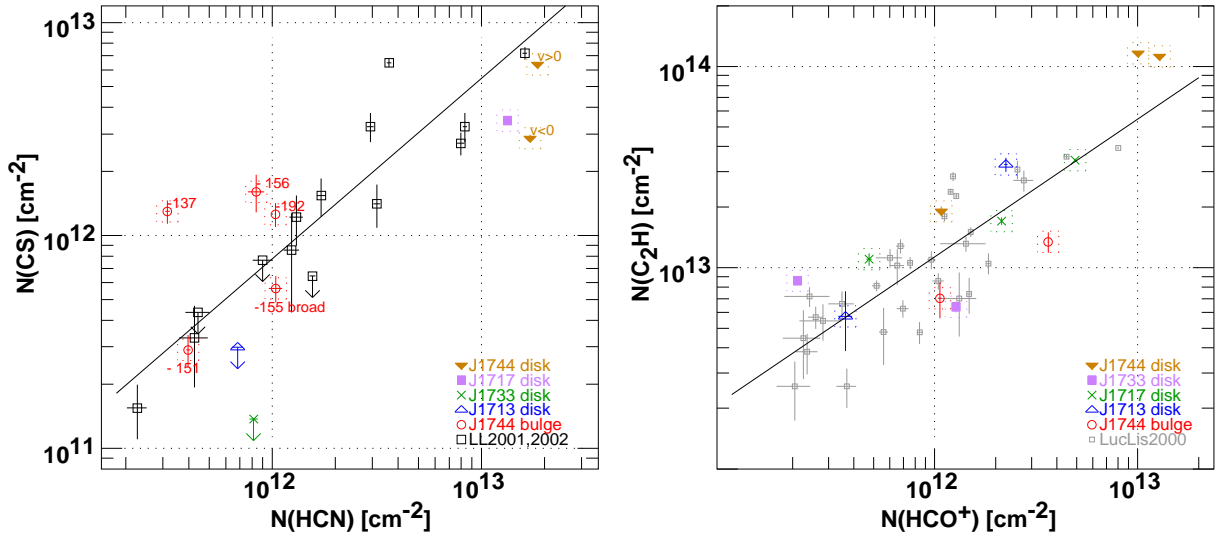}
\caption{Column densities for new and existing data. Left: N(CS) vs N(HCN).
The regression line 
is a power law of slope 0.85 fit to the existing data for clouds seen locally by 
\cite{LisLuc01} and \cite{LucLis02}.  The velocities 
of features arising in the bulge toward J1744 are indicated.  The
disk gas toward J1744 is shown separately for positive and negative velocities.
Right: N(\cch) vs N(\hcop).  The regression line 
is a power law of slope 0.70 fit to the existing data for clouds shown in Fig. 4 of
\cite{LucLis00C2H}.}
\end{figure*}


\begin{table*}
\caption[]{Observed and derived quantities}
{
\small
\begin{tabular}{lccccccccccc}
\hline
Source&l,b&S$_\nu$&${{\rm R}_<}^a$&{A$_{\rm K}$}$^b$&{N(H I)$_0$}$^c$&{N(H I)$_\tau$}$^d$&2N(\HH)$^e$&N(H)&\fH2$^f$&N(H)/A$_{\rm K}$\\
&\degr,\degr&Jy &kpc &mag &$10^{22}\pcc$&$10^{22}\pcc$&$10^{22}\pcc$&$10^{22}\pcc$&&$10^{22}\pcc$mag$^{-1}$ \\
\hline
J1713-3418 &$ 351.705,+2.787$& 0.93  & 1.23& 0.40 & 0.43 & (0.64)$^g$ & 0.18 & (0.82) &(0.22)  & $>$1.53 \\
J1717-3341 &$ 352.733,+2.391$& 0.49  & 1.08 & 0.66 & 0.54 & 0.78 & 0.53 & 1.31 & 0.40 & 1.98 \\ 
J1733-3722 &$ 351.818,-2.889$ & 0.60 &1.21 & 0.60  & 0.56  &(1.13) & 0.11 & (1.24)  &  (0.09) & $>$1.12  \\
J1744-3116 &$357.863,-0.997$ & 0.33 &0.32 & 1.46 & 0.90 & 1.20 & 1.93 & 3.13 & 0.62 & 2.14 \\
\hline
\end{tabular}}
\\
$^a$R$_< = {\rm R}_0\sin(|l|)$, R$_0= 8500$ pc  \\
$^b$values from \cite{SchChe+14}; E(B-V)/A$_{\rm K} = 3.09$ \citep{SchFin11}\\
$^c$N(H I)$_0=1.823\times 10^{18}\pcc \int T_B({\rm v}) {\rm dv}$ \\
$^d$N(H I)$_\tau$ from Eq. 1 \\
$^e$N(\HH) = N(\hcop)/$3\times 10^{-9}$, N(\hcop) $= 1.10\times 10^{12}\pcc \int \tau_\hcop {\rm dv}$ \\
$^f$\fH2\ = 2N(\HH)/(N(H I) + 2N(\HH)) \\
$^g$Parenthesized quantities assume N(H)/A$_{\rm K}$ = $2.06\times 10^{22}\pcc$ mag$^{-1}$, the average for  J1717 and J1744 \\
\end{table*}


\begin{table*}
\caption[]{$^{12}$C/$^{13}$C isotopic abundance ratios from \hcop}
{
\small
\begin{tabular}{lccccc}
\hline
Source & velocity range & W$_\hcop$ & W$_{{\rm H}^{13}{\rm CO}^+}$ &N(\hcop)$^a$ & N(\hcop)/N(H$^{13}$CO\p)$^a$ \\
       & \kms           & \kms      & \kms & $10^{12}\pcc$ &  \\
\hline
J1744 & -200..-180 & 0.890(0.014) & (0.017) & 0.98(0.02)& $>17^b$ \\
J1744 & -165..-135 & 3.301(0.019) & (0.021) & 3.63(0.02)& $>51$ \\
J1744 & -200..-135 & 4.191(0.024) & (0.027) & 4.68(0.03)& $>51$ \\
\hline
J1713 & -38..14 & 2.321(0.053) & (0.046) & 2.55(0.06) &$>16$ \\
J1717 & -25..10 & 7.002(0.027) & 0.117(0.018) & 7.70(0.03)&$58\pm 9$ \\
J1744 & -33..13 & 21.852(0.056) & 0.333(0.023) &24.03(0.06) & $64\pm 4$ \\
\hline
\end{tabular}}
\\
$^a$ N(\hcop)$ = 1.10\times10^{12}\pcc$ W$_\hcop$; 
N(H$^{13}$CO\p) $ = 1.135\times10^{12}\pcc$ W$_{{\rm H}^{13}{\rm CO}^+}$ \\
$^b$ all limits are 3$\sigma$ \\
\end{table*}

\begin{table}
\caption[]{Gaussian decomposition of disk H I absorption toward J1744}
{
\small
\begin{tabular}{lcccc}
\hline
comp  & center & $\tau_0$ & FWHM & {W$_{\rm H I}$}$^a$ \\    
      & \kms   &          & \kms & \kms \\
\hline
  1 & -22.602  &   1.515  &  5.876 & 9.4763 \\
$\pm$ &  0.127  & 0.109  &  0.260 &  0.515 \\
  2  & -12.476  &   0.646  & 3.603  & 2.477 \\
$\pm$  & 0.154  & 0.0928  & 0.411  & 0.290 \\
  3  & 0.127  &   0.967  & 17.369  & 17.874 \\
$\pm$  &  0.579 &  0.0621  &  0.957  &  0.976 \\
  4  & 7.091  &   0.686  & 4.587  &  3.348 \\
$\pm$  & 0.181 &  0.081  & 0.442  & 0.330 \\
\hline
\end{tabular}} 
\\
\end{table}

\begin{table}
\caption[]{Gaussian decomposition of bulge CS absorption}
{
\small
\begin{tabular}{lcccc}
\hline
comp  & center & $\tau_0$ & FWHM & {W$_{\rm CS}$}$^a$ \\    
      & \kms   &          & \kms & \kms \\
\hline  
1 & -192.6734 &   0.1440 & 1.0171 &  0.1559 \\
($\pm$)& 0.0184 &  0.0056 & 0.0426 & 0.0057 \\
  2 & -156.6340 &   0.0633 & 0.6582 &  0.0443 \\
($\pm$)&  0.0332 & 0.0065 & 0.0812 & 0.0046 \\
  3 & -155.5189 &   0.3153 & 0.5920 & 0.1987 \\
($\pm$)&  0.0072 & 0.0085&  0.0167 & 0.0050 \\
  4 & -136.7139 &   0.2349 & 0.6433 & 0.1609 \\
($\pm$) &  0.0094 & 0.0076 & 0.0215 & 0.0048 \\
\hline
\end{tabular}}
\\
$^a$N(CS)$=8.06\times10^{12}\pcc~{\rm W}_{\rm CS}$ \\
\end{table}

\begin{table}
\caption[]{Gaussian decomposition of bulge HCN absorption}
{
\small
\begin{tabular}{lcccc}
\hline
comp  & center & $\tau_0$ & FWHM & {W$_{\rm HCN}$}$^a$ \\    
      & \kms   &          & \kms & \kms \\
\hline
  1 &  -192.6652 &   0.3406 & 0.8340 &   0.5480 \\
$(\pm)$ &  0.0069 & 0.0037 &  0.0146 &  0.0040 \\ 
  2 & -155.8931 &   0.2841 &  0.8530 &  0.4643  \\
$(\pm)$ &  0.0089 &  0.0043 & 0.0208 &  0.0048  \\
  3 & -155.5423 &  0.0297 &  8.8886 &  0.5058  \\
$(\pm)$ &  0.2969 &  0.0027 &  1.4085  & 0.0331  \\
  4 & -151.8292 &   0.1090 &  0.9935  &  0.2075  \\
$(\pm)$ &  0.0245 &  0.0040 & 0.0616  &  0.0054  \\
  5 & -137.2511 &   0.1070 &  0.8153  & 0.1672  \\
$(\pm)$ &  0.0200&  0.0046&  0.0471  & 0.0043   \\
\hline
\end{tabular}} 
\\
$^a$N(HCN)$=1.89\times10^{12}\pcc~{\rm W}_{\rm HCN}$ \\
\end{table}

\begin{table}
\caption[]{\hcop\ and \cch\ optical depth integrals in bulge and disk gas}
{
\small
\begin{tabular}{lccc}
\hline
Source & velocity range & {W$_\hcop$}$^a$ & {W$_{{\rm C}_2{\rm H}}$}$^b$ \\
       & \kms           & \kms       & \kms   \\
\hline
J1744 & -210..-180 & 0.964(0.018) & 0.108(0.022) \\
J1744 & -165..-130 & 3.296(0.019) & 0.206(0.024) \\
\hline
J1713 & -40..-22 & 0.330(0.026) & 0.088(0.026) \\
J1713 & -21..14 &2.038(0.038) & 0.499(0.040) \\
J1717 & -27..-10 & 0.432(0.013) & 0.169(0.012) \\
J1717 & -8..0 & 4.640(0.021) & 0.524(0.009) \\
J1717 & 0..11 &1.942(0.013) & 0.261(0.011) \\
J1733 & -3..6 &1.156(0.008) & 0.098(0.006) \\ 
J1733 & 6..11 & 0.592(0.005) & 0.132(0.005) \\
J1744 & -31..-20 & 0.978(0.015) & 0.292(0.016) \\
J1744 & -16..0 & 9.144(0.046) & 1.784(0.020) \\
J1744 & 0..15 & 11.637(0.052) & 1.719(0.020) \\
\hline
\end{tabular}}
\\
$^a$ N(\hcop)$ = 1.10\times10^{12}\pcc$ W$_\hcop$ \\
$^b$ For the strongest hyperfine component only \\
  N(\cch)$ = 6.52\times10^{13}\pcc$ W$_{{\rm C}_2{\rm H}}$ \\
\end{table}

\section{Summary}

Following up on the serendipitous detection of molecular absorption 
at velocities characteristic of the Galactic bulge toward J1744-3116 
($l=-2.13$\degr, $b = -1.00$\degr), we observed molecular absorption 
toward three other quasars seen against the bulge in the fourth longitude 
quadrant, although at much larger longitude
separation from the center and higher Galactic latitude (Table 1).  We
also reobserved the absorption toward J1744 with higher spectral
resolution and broader spectral coverage, to better characterize the 
bulge gas. The species and transitions observed were \hcop\ (1-0), HCN(1-0), 
\cch (1-0), CS(2-1) and H$^{13}$CO\p (1-0), all in the frequency range 87 - 98 GHz.
Line profiles are shown in Figs. 1-3 and general properties of the 
sightlines are summarized in Table 1.

\subsection{Absorption arising inside the Galactic disk}

The 3 kpc arm appeared in \hcop\ absorption at -90 \kms\ toward J1717-3341 
and we inferred a molecular hydrogen fraction  0.24.  We also detected a 
weak molecular absorption feature at $+52$ \kms\ toward J1744 from the 
far-side 3 kpc arm and inferred a higher molecular fraction, 0.48, in this
case.  No molecular absorption
was observed from the bulge gas at -200 \kms\ $\la$ v $\la$  -140 \kms\
toward J1717, corresponding to the H I emission that was modelled as part 
of the tilted H I inner-galaxy gas distribution by \cite{BurLis78}.

Absorption from CS and HCN in the bulge gas toward J1744 consists of several 
narrow lines with FWHM of 1 \kms\ or less (Tables 4-5 and Fig. 5). In \hcop\ the 
narrow lines seen in CS and HCN (and other lines not seen in CS or HCN) are 
superimposed on a broader, more continuous absorption, indicating that 
\hcop\ is more widespread and more broadly distributed in space, as is the case 
with diffuse molecular gas observed near the Sun and more generally
in the Galactic disk.  The appearance of narrow lines is untypical of
gas that is observed inside the Galactic disk, but the proliferation
of so many features, and sub-structure in the \hcop\ profile, may be
ascribed to the very large line of sight velocity gradient due to
the viewing geometry.  This can cause a velocity separation of gas 
parcels that are relatively close, and have little relative motion with 
respect to each other.

 Comparing CS with HCN and \cch\ with \hcop\ (Fig. 6) we showed that 
their chemistries in the bulge gas observed toward J1744 at 
v $<$ -135 \kms\ resemble those seen in absorption in local and 
disk  diffuse molecular gas.  We set a limit N(\hcop)/N(H$^{13}$CO\p)
$>$ 51 (3$\sigma$) for the bulge gas toward J1744 (Table 2), showing that it 
is not enriched in $^{13}$C as is material in the central molecular zone 
within 200 pc of Sgr A*. The bulge gas seen toward J1744 just outside the
CMZ is very clearly differentiated from gas in the CMZ, and 
similar to disk gas in every aspect we examined.

\subsection{Molecular gas in the Galactic disk}

All of the sightlines showed molecular absorption from disk gas at 
$|\rmv| \la 40$ \kms\ in \hcop, \cch\ and HCN, and two showed absorption 
in disk gas from CS (2-1).  Toward J1717 and J1744 we measured isotopologic 
abundance ratios N(\hcop)/N(H${13}$CO\p) = 58$\pm9$ and $64\pm4$, 
respectively. These are smaller than the Solar abundance ratio 89, but 
comparable to values seen previously in $\lambda3$mm absorption from local 
diffuse molecular gas.  Chemical abundances in the disk gas are shown 
in Fig. 6: they are generally like that seen previously for diffuse 
molecular gas seen at high Galactic latitude near the Sun, with some 
peculiarities: high \cch/\hcop\ ratios toward J1744 and small CS/HCN 
ratios toward J1733 and at v $<$ 0 \kms\ toward J1744.

\begin{acknowledgements}

This paper makes use of the following ALMA data: ADS/JAO.ALMA\#2016.1.00132.S .
ALMA is a partnership of ESO (representing its member states), NSF (USA)
and NINS (Japan), together with NRC (Canada), NSC and ASIAA (Taiwan), and
KASI (Republic of Korea), in cooperation with the Republic of Chile.  The
Joint ALMA Observatory is operated by ESO, AUI/NRAO and NAOJ.
The National Radio Astronomy Observatory is a facility of the National Science 
Foundation operated under cooperative agreement by Associated Universities, Inc. 

This work was supported by the French program “Physique et Chimie du Milieu 
Interstellaire” (PCMI) funded by the Conseil National de la Recherche Scientifique 
(CNRS) and Centre National d’Etudes Spatiales (CNES).  HSL is grateful to the
hospitality of the ITU-R, the Hotel Bel Esperance in Geneva, the JAO in
Santiago and LHotel in Montreal during the completion of this manuscript. 
We thank Chinshin Chang at the ALMA JAO for producing the imaging scripts 
for this project and sheparding it through the data reduction process and we
thank the referee for a close reading of the manuscript.

\end{acknowledgements}

\bibliographystyle{apj}

\end{document}